\documentstyle[epsfig]{aipproc}

\def\to{\rightarrow}
\def\sn{\tilde\nu}
\def\mupmum{\mu^+\mu^-}
\def\cnone{\chi^0}
\def\mcnone{m_{\cnone}}

\def\snt{\tilde\nu_\tau}

\def\msnt{m_{\snt}}
\def\dm{\Delta\msnt}

\def\gamsnt{\Gamma_{\snt}}
\def\gev{~{\rm GeV}}
\def\sigsntbar{\overline \sigma_{\snt}}
\def\rp{\not{\hbox{\kern-4pt $R_P$}}}
\def\met{\not{\hbox{\kern-4pt $E_T$}}}
\def\mpt{\not{\hbox{\kern-4pt $p_T$}}}
\def\fbi{\text{ fb}^{-1}}
\def\anti{\overline}
\def\rts{\protect\sqrt{s}}
\def\sigrts{\sigma_{\!\rts}}

\begin{document}

\noindent
\begin{minipage}[t]{\textwidth}
\begin{flushright}
hep-ph/9801248 \hfill
LBNL--41253 \\
January 1998 \hfill
UCB--PTH--98/03 \\
\end{flushright}
\end{minipage}
\vspace*{.3in}

\title{
$\bbox{R}$-Parity Violation and Sneutrino Resonances at Muon Colliders
\thanks{This work was supported in part by the Director, 
Office of Energy Research, Office of High Energy and Nuclear Physics,
Division of High Energy Physics of the DOE under Contracts
DE--AC03--76SF00098 and by the NSF under grant PHY--95--14797.}  }

\author{Jonathan L. Feng}

\address{Theoretical Physics Group, Lawrence Berkeley
National Laboratory \\ and Department of Physics, University of
California, Berkeley, CA 94720}

\maketitle

\begin{abstract}
\normalsize
In supersymmetric models with $R$-parity violation, sneutrinos may be
produced as $s$-channel resonances at $\mu^+ \mu^-$ colliders.  We
demonstrate that, for $R$-parity violating couplings as low as
$10^{-4}$, sneutrino resonances may be observed and may be exploited
to yield high precision SUSY parameter measurements.  The excellent
beam energy resolution of muon colliders may also be used to resolve
MeV level splittings between CP-even and CP-odd sneutrino mass
eigenstates.

\small
\vspace*{0.6in}
\begin{center}
To appear in\\
\vspace*{0.05in}
Proceedings of the Workshop on Physics at the First Muon Collider\\
and at the Front End of a Muon Collider \\
Fermi National Accelerator Laboratory, 6--9 November 1997 \\
\vspace*{0.05in}
and\\
\vspace*{0.05in}
Proceedings of the 4th International Conference on\\
Physics Potential and Development of $\mu^+ \mu^-$ Colliders\\
San Francisco, 10--12 December 1997
\end{center}

\end{abstract}

\newpage
\pagestyle{plain}

Low-energy supersymmetry (SUSY) is a leading candidate for physics
beyond the standard model (SM).  When exploring the physics
opportunities at a muon collider, it is therefore important to
consider its potential for discovering supersymmetric particles and
determining SUSY parameters.

For many studies, the muon collider's potential parallels that of more
extensively studied $e^+e^-$ colliders, with obvious modifications for
differences in luminosity and beam polarization.  In fact, as studies
of LEP II typically assume $\sqrt{s} \sim 190 \gev$ and a total
integrated luminosity of $\sim 1\fbi$, characteristics similar to
those of the proposed First Muon Collider (FMC), many interesting
results from these studies apply equally well to the FMC.  For
example, from chargino production at LEP II or the FMC, gaugino mass
unification and the viability of the LSP as a dark matter candidate
can be tested in a highly model-dependent manner~\cite{FS}.

There are, however, essential differences that warrant more careful
study. Most obviously, if a muon collider reaches $\sqrt{s} \sim 4$
TeV, a great number of complicated sparticle signals may be
present~\cite{GK}, as well as a number of backgrounds that have not
yet been intensively studied.  In addition, the excellent beam energy
resolution of muon colliders is promising for precise mass
measurements, whether through threshold scans~\cite{Berger} or
kinematic endpoints~\cite{Lykken}.

In this study, we will consider what a muon collider may bring to the
study of $R$-parity violating ($\rp$) SUSY theories.  When $R$-parity
is violated, the distinction between neutral Higgs bosons and scalar
neutrinos is blurred, and so scalar neutrinos are also produced as
$s$-channel resonances at lepton colliders.  As with Higgs resonances,
such resonances may be highly suppressed at electron colliders.  At
muon colliders, however, we will see that even if $\rp$ couplings are
comparable to their Yukawa coupling counterparts, sneutrino resonances
may be exploited to yield high precision measurements of SUSY
parameters.  Further details may be found in Ref.~\cite{FGH}.  (See
also Ref.~\cite{ray}.)

$R$-parity is defined to be $R_P =+1$ and $-1$ for SM particles and
their superpartners, respectively.  If $R$-parity is conserved, all
superpartners must be produced in pairs.  However, renormalizable
gauge-invariant interactions that explicitly violate $R$-parity and
lepton number are also allowed by the superpotential

\begin{eqnarray}
W &=& \lambda L L E^c + \lambda' L Q D^c \nonumber \\
  &=& \lambda_{ijk} ( N_i E_j E^c_k - E_i N_j E^c_k )
+ \lambda'_{lmn} ( N_l D_m D^c_n - V^*_{pm} E_l U_p D^c_n)\ ,
\label{superpotential}
\end{eqnarray}
where the lepton and quark chiral superfields $L = (N, E)$, $E^c$, $Q
= (U, D)$, and $D^c$ contain the SM fermions $f$ and their scalar
partners $\tilde{f}$, $V$ is the CKM matrix, $i<j$, and all other
generational indices are arbitrary.  With the couplings of
Eq.~(\ref{superpotential}), superpartners may be produced singly at
colliders.  In particular, sneutrinos $\sn$ may be produced as
$s$-channel resonances at lepton colliders through the $\lambda$
couplings~\cite{snu,efp}.  Such resonance production is unique in that
it probes supersymmetric masses up to $\sqrt{s}$. As sneutrinos are
likely to be among the lighter superparticles, even a first stage muon
collider with $\sqrt{s} = 80 - 250 \text{ GeV}$ will cover much of the
typically expected mass range.  We will explore the potential of a
muon collider to study such resonances, assuming luminosity and beam
resolution options $({\cal L}, R) = (1\ {\rm fb}^{-1}{\rm /yr},
0.1\%)$ and $(0.1\ {\rm fb}^{-1}{\rm /yr}, 0.003\%)$.

\begin{figure}[t]
\centerline{\epsfig{file=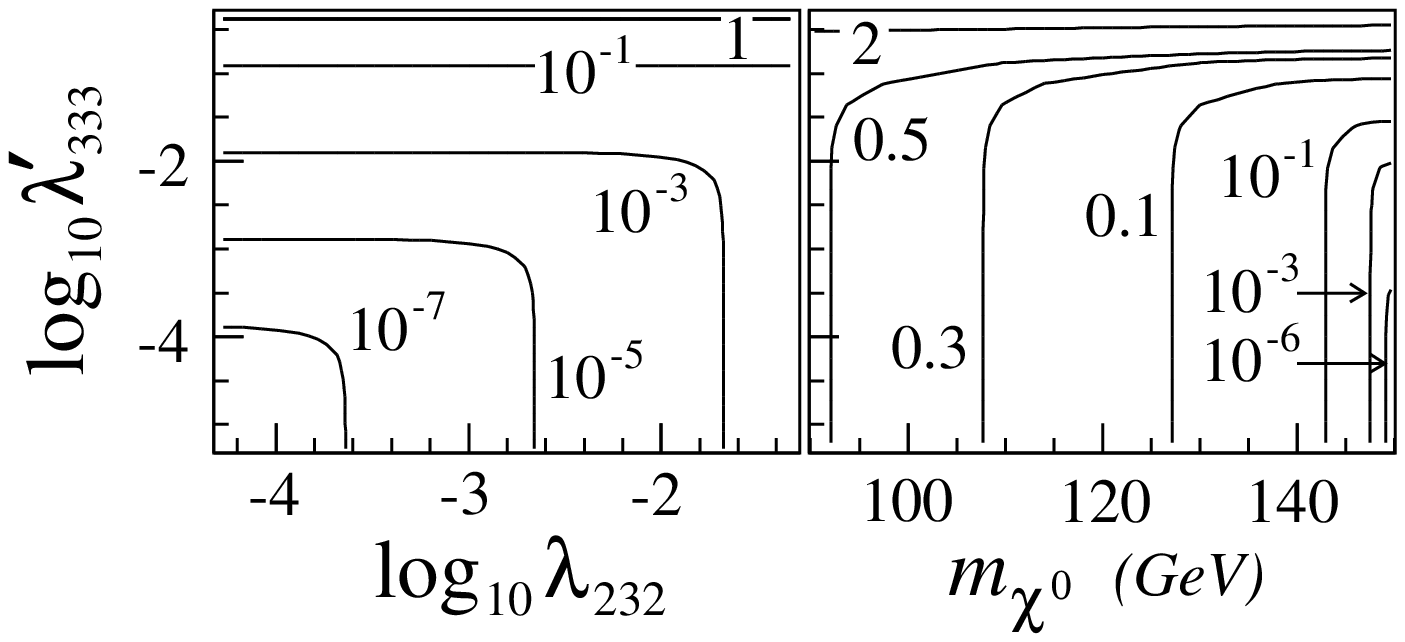,width=0.89\textwidth}}
\vspace*{0.1in}
\caption{Contours of total decay width $\gamsnt$ in GeV (i) for
$\msnt = 100 \gev$, assuming only $\rp$ decays $\snt \to \mu^+\mu^-,
b\anti{b}$ are open, and (ii) for $\msnt = 150\gev$, assuming that
$\snt \to \nu_{\tau} \cnone$ decays are also allowed, with $\cnone =
\tilde{B}$ and fixed $\lambda_{232}=5\times 10^{-5}$.}
\label{fig:widths}
\end{figure}

At muon colliders, sneutrinos $\sn_e$ and $\sn_{\tau}$ may be produced
in the $s$-channel.  They can then decay through $\lambda$
($\lambda'$) couplings to charged lepton (down-type quark) pairs or
through $R_P$-conserving decays, such as $\sn \to \nu
\cnone$.  In the latter case, the lightest neutralino $\cnone$ 
subsequently decays to three SM fermions through $\rp$ interactions.
The phenomenology of sneutrino resonances is thus rather complicated
in full generality.  However, in analogy with the Yukawa couplings,
$\rp$ couplings involving higher generational indices are usually
expected to be larger.  We therefore focus on $\sn_\tau$ production
through the coupling $\lambda_{232}$, and, in addition to the decay
$\sn_\tau \to \mu^+ \mu^-$, consider the possibility of $\sn_\tau \to
b \anti{b}$ decays governed by $\lambda'_{333}$.  For simplicity, we
take these two $\rp$ couplings to be real and assume that all other
$\rp$ parameters are negligible.  The current bounds on these
couplings arising from a variety of sources~\cite{FGH} are
$\lambda_{232} \lesssim 0.06$, $\lambda'_{333} \lesssim 1$, and
$\lambda_{232} \lambda'_{333} \lesssim 0.001$.  We will also consider
a scenario in which the $R_P$-conserving decay $\snt \to \nu_{\tau}
\cnone$ is important.  Fig.~\ref{fig:widths} shows representative
decay widths for the three modes.

The cross section for resonant $\snt$ production is

\begin{equation}
\sigma_{\snt}(\sqrt{s}) = {8\pi\Gamma(\snt\to\mu^+\mu^-)
\Gamma(\snt\to X)\over (s-\msnt^2)^2 +\msnt^2\gamsnt^2} \ ,
\label{resonancexsec}
\end{equation}
where a factor of 2 has been explicitly included to account for both
$\snt$ and $\snt^*$ exchange, $X$ denotes a generic final state from
$\snt$ decay, and $\gamsnt$ is the total sneutrino decay width.  The
effective cross section $\sigsntbar$ is obtained by convoluting
$\sigma_{\snt}(\rts)$ with the collider's $\rts$ distribution.
Neglecting (for purposes of discussion) bremsstrahlung and
beamstrahlung, this distribution is well-approximated by a Gaussian
distribution with rms width $\sigrts = 7~{\rm MeV}\left[R/
0.01\%\right]\left[\sqrt s/{\rm 100\ GeV}\right]$, where $R$ is the
beam energy resolution factor.  In two extreme limits, $\sigsntbar$
can be expressed in terms of branching fractions $B$ as

\begin{eqnarray}
\gamsnt \ll \sigrts \, &:&\ \ \sigsntbar (\msnt) \simeq
\frac{\sqrt{8\pi^3}}{\msnt^2} \frac{\gamsnt}{\sigrts}
B(\mu^+ \mu^- ) B(X) \, , \nonumber \\
\gamsnt \gg \sigrts \, &:&\ \ \sigsntbar (\msnt) \simeq 
\frac{8\pi}{\msnt^2} B(\mu^+ \mu^- ) B(X) \ .   
\label{widthxsec}
\end{eqnarray}
If only highly suppressed $\rp$ decays are present, $\sigsntbar
\propto \gamsnt/\sigrts$.  The small values of $\sigrts$ possible 
at a muon collider thus provide an important advantage for probing
small $\rp$ couplings. At a muon collider, the effects of
bremsstrahlung are small (but are included in our numerical results);
beamstrahlung is negligible.

The signals for $\snt$ production depend on the $\snt$ decay
patterns. We consider two well-motivated scenarios. In the first,
$\msnt < \mcnone$, and $\snt$ decays only through $\rp$
operators. Neglecting $\rp$ couplings other than $\lambda_{232}$ and
$\lambda'_{333}$, the signal is $\mu^+ \mu^-$ or $b\anti{b}$ pairs in
the final state.  For concreteness, we consider $\msnt = 100\gev$.

The dominant backgrounds are Bhabha scattering and $\mu^+\mu^- \to
\gamma^*, Z^* \to \mu^+\mu^-, b\anti{b}$.  To reduce these, we apply 
the following cuts: for the $\mu^+\mu^-$ ($b\anti b$) channel, we
require $60^{\circ} <\theta < 120^{\circ}$ ($10^{\circ} < \theta <
170^{\circ}$) for each muon ($b$ quark). The stronger $\theta$ cuts in
the $\mupmum$ channel are needed to remove the forward-peaked Bhabha
scattering.  We also require $|m_{f\anti{f}} - \msnt | < 7.5\gev$ in
both channels to reduce background from radiative returns to the $Z$.
After the cuts above and including beam energy spread and
bremsstrahlung, the background cross sections at $\sqrt{s} = 100\gev$
are $\sigma(\mupmum)=3.5\times 10^4~{\rm fb}$ and $\sigma(b\anti
b)=2.0\times 10^5~{\rm fb}$.

\begin{figure}[t]
\centerline{\epsfig{file=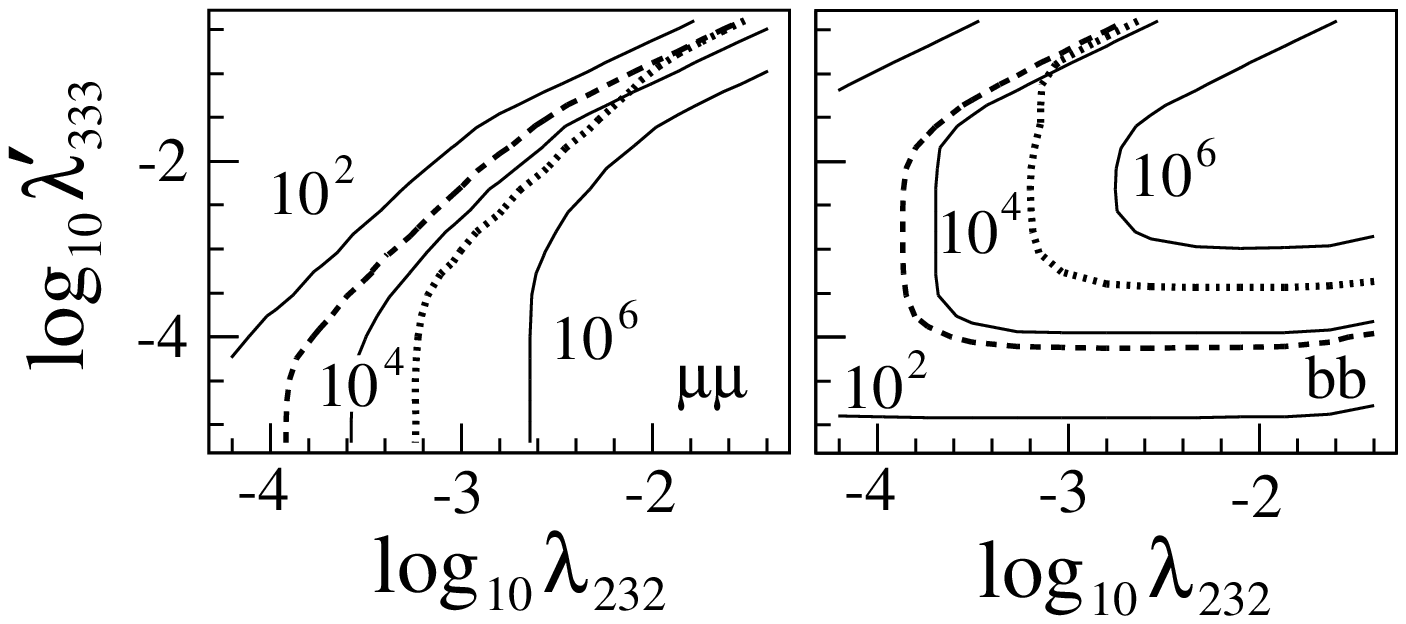,width=0.89\textwidth}}
\vspace*{0.1in}
\caption{Contours for (i) $\sigma(\mu^+ \mu^- \to
\snt \to \mu^+ \mu^- )$ and (ii) $\sigma(\mu^+ \mu^- \to
\snt \to b\anti{b})$ (solid) in fb after cuts for the $\msnt 
< m_{\cnone}$ scenario, with $\protect\sqrt{s} = \msnt= 100\gev$ and
$R=0.003\%$.  The dashed and dotted contours give the optimistic and
pessimistic/scan $3\sigma$ discovery boundaries, respectively, for
total integrated luminosity $L = 0.1\fbi$.  (See discussion in text.)}
\label{rates1}
\end{figure}

In this scenario, $\gamsnt$ is unknown {\em a priori}, but a very
small $\gamsnt$ is possible.  We choose the $({\cal L},R)=
(0.1\fbi/\text{yr}, 0.003\%)$ option, which maximizes $S/\sqrt B$ if
$\gamsnt$ is indeed small. With this choice, signal cross sections
after cuts are given by the solid contours in Fig.~\ref{rates1}. We
see that the cross sections may be extremely large ($>1$ nb) in some
regions of the allowed parameter space.

In Fig.~\ref{rates1} we also give sneutrino resonance discovery
contours for two extreme possibilities. In the most optimistic case,
the sneutrino mass is exactly known and the total luminosity is
applied at the sneutrino resonance peak. The corresponding
``optimistic'' $3\sigma$ discovery contours are given by dashed
lines. (In calculating $S/\sqrt{B}$ for the $b\anti{b}$ mode here and
below, we include a 75\% efficiency for tagging at least one $b$
quark.) More realistically, the sneutrino mass will be known only
approximately from other colliders with some uncertainty $\pm
\frac{1}{2} \dm$; we assume $\dm = 100\text{ MeV}$ using the fully
reconstructable $\rp$ decays.  The dotted contours of
Fig.~\ref{rates1} represent the 3$\sigma$ ``pessimistic/scan'' $\snt$
discovery boundaries, where the effects of having to scan over the
allowed sneutrino mass interval are included. (See Ref.~\cite{FGH} for
details.) The actual discovery limit will lie between the dashed and
dotted contours.  We see that $\snt$ resonance observation is possible
for $\rp$ couplings as low as $10^{-3} - 10^{-4}$.

We now consider a second scenario in which $m_{\sn_\tau} >
m_{\cnone}$.  In addition to $\rp$ decays, decays $\snt\to\nu_{\tau}
\cnone$ are now also allowed and typically dominate, with $\cnone$
then decaying to $\nu_{\tau}\mu\mu$ or $\nu_{\mu} \mu \tau$ through
the $\lambda_{232}$ coupling, or $\nu_{\tau}b\anti{b}$ through the
$\lambda'_{333}$ coupling.  The final signals are then $\mupmum +
\met$, $\mu^{\pm}\tau^{\mp} + \met$, and $b\anti{b} + \met$. For this
scenario, we consider masses $\msnt = 150\gev$ and $\mcnone =
100\gev$.

The leading backgrounds to the $\nu\cnone$ channels are from
$WW^{(*)}$ and $ZZ^{(*)}$. To reduce these, we require $\met > 25
\text{ GeV}$, that the visible final state fermions have $p_T >
25\gev$ and $60^{\circ} < \theta < 120^{\circ}$ for the lepton modes
($40^{\circ} < \theta < 140^{\circ}$ for the $b\anti{b}\met$ mode),
and that the invariant mass of the two visible fermions be $> 50\gev$.
With these cuts, the total combined background in the $\nu\cnone$
channels is $\sim 1$ fb.

\begin{figure}[t]
\centerline{\epsfig{file=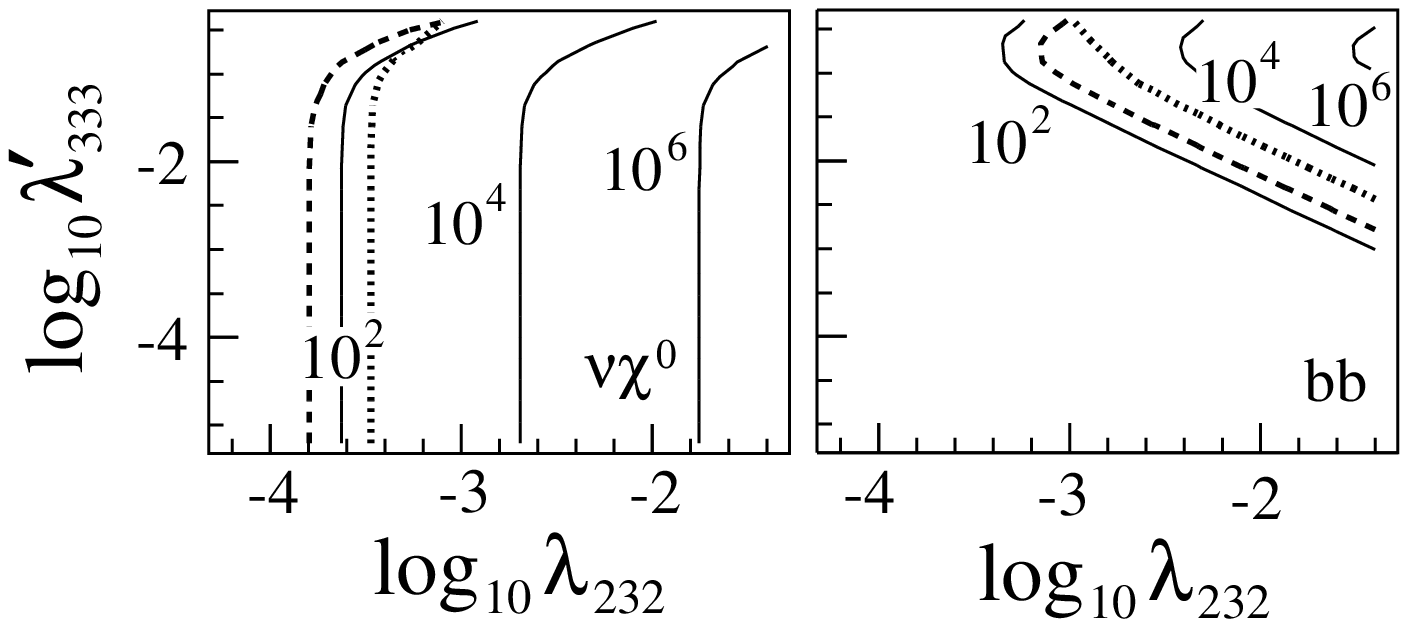,width=0.89\textwidth}}
\vspace*{0.1in}
\caption{Contours for (i) $\sigma(\mu^+ \mu^- \to \snt \to \nu 
\cnone)$ (no cuts) and (ii) $\sigma(\mu^+ \mu^- \to \snt \to b 
\anti{b})$ (after cuts) in fb assuming $\msnt = 150\gev$, 
$\mcnone = 100\gev$, and $\cnone= \tilde{B}$.  The optimistic (dashed)
and pessimistic/scan (dotted) discovery contours assume $L=1\fbi$ and
$R=0.1\%$.}
\label{rates2}
\end{figure}

The signal cross sections for the $\nu\cnone$ channel (without cuts)
and the direct $\rp$ $b\anti{b}$ channel (after cuts as in
Fig.~\ref{rates1}) are plotted in Fig.~\ref{rates2}.  We also give
3$\sigma$ discovery contours for both optimistic and pessimistic/scan
cases as before, where we choose the $({\cal L},R)= (1\fbi/\text{yr},
0.1\%)$ option to maximize $\cal L$. For the ``pessimistic/scan''
discovery contours, we assume $\dm\sim 2\gev$ from kinematic
endpoints.  We see that the nearly background-free $\nu\cnone$ mode
makes possible a dramatic improvement in discovery reach compared to
the $\msnt<\mcnone$ scenario. The $\snt$ resonance may be discovered
for $\lambda_{232} \gtrsim 10^{-4}$, irrespective of the value of
$\lambda'_{333}$.

\begin{figure}[t]
\centerline{\epsfig{file=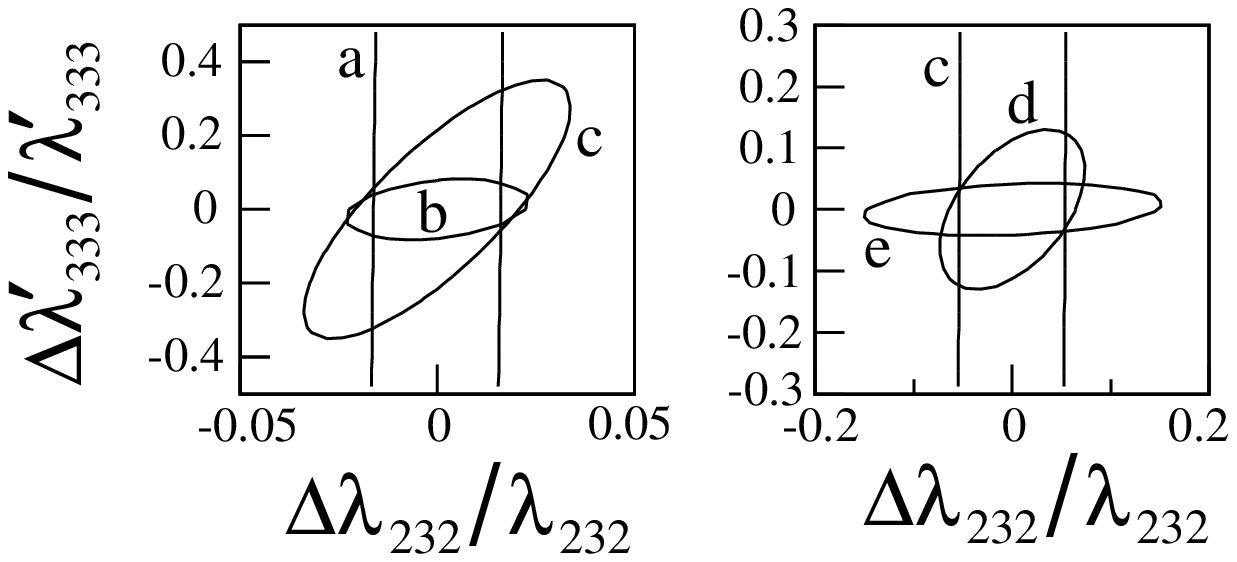,width=0.89\textwidth}}
\vspace*{0.1in}
\caption{$\chi^2 = 1$ contours in the $(\Delta\lambda_{232} 
/ \lambda_{232}, \Delta\lambda'_{333} / \lambda'_{333})$ plane for (i)
the $\msnt = 100 \gev < \mcnone$ scenario, assuming $L=0.3\fbi$,
$R=0.003\%$ and (ii) the $\msnt = 150\gev$ $>$ $\mcnone = 100\gev$
scenario, assuming $L=3\fbi$, $R=0.1\%$.  Contours are for
$\lambda_{232}=5\times 10^{-4}$ and $\lambda'_{333}=$: (a) $10^{-5}$;
(b) $5\times 10^{-4}$; (c) $10^{-2}$; (d) $10^{-1}$; (e) $0.3$.}
\label{signif}
\end{figure}

Once we have found the sneutrino resonance via the scan described, the
crucial goal will be to precisely measure the relevant $\rp$
couplings. In the $\msnt<\mcnone$ scenario, the discovery scan gives a
precise determination of $\msnt$ (and, if $\gamsnt>2\sigrts$, a rough
determination of $\gamsnt$). We then envision accumulating $L=0.1\fbi$
($R=0.003\%$) at each of the three points $\rts=\msnt$, $\msnt\pm
\Delta\rts/2$, where $\Delta \rts=\text{max}[2\sigrts,\gamsnt]$.  
The off-resonance points ensure good sensitivity to $\gamsnt$.  This
is especially crucial when $\gamsnt>\sigrts$, as in this case a single
measurement of $\sigsntbar$ at $\rts=\msnt$ determines
$B(\snt\to\mupmum)$ but not $\Gamma(\snt\to\mupmum)$; see
Eq.~(\ref{widthxsec}).  In the $\msnt>\mcnone$ scenario, we noted that
$\gamsnt$ can be computed with good precision from observations at
other colliders; we assume a $\pm 5\%$ error for $\gamsnt$. We would
then run only at $\rts\simeq \msnt$ and accumulate $L=3\fbi$
($R=0.1\%$). In Fig.~\ref{signif}, the resulting $\chi^2 = 1$ error
contours are plotted for each of the two scenarios. We find that
$1\sigma$ fractional errors at the few percent level can be achieved,
even for a small value of $\lambda_{232}=5\times 10^{-4}$, which is
not very far inside the discovery regions.

As a final remark, we note that $\rp$ interactions can split the
complex scalar $\snt$ into a real CP-even and a real CP-odd mass
eigenstate.  This splitting is generated both at tree-level (from
sneutrino-Higgs mixing) and radiatively, and both contributions depend
on many SUSY parameters.  However, such $\rp$ terms also generate
neutrino masses, and it is generally true that the sneutrino
splittings generated are ${\cal O}(m_{\nu})$~\cite{g}.  Given the
current bound $m_{\nu_{\tau}} < 18.2 \text{ MeV}$\cite{nutaumass}, we
see that $\tau$ sneutrino splittings may be as large as ${\cal O} (10
\text{ MeV})$. A muon collider with $R=0.003\%$ is uniquely capable of
resolving resonance peak splittings at or below the MeV level.

In summary, we have demonstrated that a muon collider is an excellent
tool for discovering sneutrino resonances and measuring their
$R$-parity violating couplings.  Note that for small $\rp$ couplings,
absolute measurements through other processes and at other colliders
are extremely difficult, as they typically require that $\rp$ effects
be competitive with a calculable $R_P$-conserving process. For
example, $\rp$ neutralino branching ratios constrain only ratios of
$\rp$ couplings.  In addition, a muon collider is unique in its
ability to resolve the splitting between the CP-even and CP-odd
sneutrino components when this splitting is as small as expected given
the current bounds on neutrino masses.

\vspace*{.1in}

I thank J.~Gunion and T.~Han for the collaboration upon which this
talk was based and the U.C. Davis theory group for hospitality during
the course of this work.  This work was supported in part by the
Director, Office of Energy Research, Office of High Energy and Nuclear
Physics, Division of High Energy Physics of the DOE under Contracts
DE--AC03--76SF00098 and by the NSF under grant PHY--95--14797.

\end{document}